# Democratization of Quantum Technologies

Zeki C. Seskir[1], Steven Umbrello[2*], Christopher Coenen[1], Pieter E. Vermaas[2]

[1] Karlsruhe Institute of Technology, Institute for Technology Assessment and Systems Analysis (ITAS), Karlsruhe, Germany

[2] Department of Values, Technology, & Innovation, School of Technology, Policy & Management, Delft University of Technology, Delft, The Netherlands

[*] Corresponding author: S.Umbrello@tudelft.nl

## Abstract

As quantum technologies (QT) advance, their potential impact on and relation with society has been developing into an important issue for exploration. In this paper, we investigate the topic of democratization in the context of QT, particularly quantum computing. The paper contains three main sections. First, we briefly introduce different theories of democracy (participatory, representative, and deliberative) and how the concept of democratization can be formulated with respect to whether democracy is taken as an intrinsic or instrumental value. Second, we give an overview of how the concept of democratization is used in the QT field. Democratization is mainly adopted by companies working on quantum computing and used in a very narrow understanding of the concept. Third, we explore various narratives and counter-narratives concerning democratization in QT. Finally, we explore the general efforts of democratization in QT such as different forms of access, formation of grassroot communities and special interest groups, the emerging culture of manifesto writing, and how these can be located within the different theories of democracy. In conclusion, we argue that although the ongoing efforts in the democratization of QT are necessary steps towards the democratization of this set of emerging technologies, they should not be accepted as sufficient to argue that QT is a democratized field. We argue that more reflexivity and responsiveness regarding the narratives and actions adopted by the actors in the QT field, and making the underlying assumptions of ongoing efforts on democratization of QT explicit, can result in a better technology for society.

## 1. Introduction

Over the course of the last several decades, the impacts of transformative technologies have been the subject of intense scholarly and public debate and scrutiny. Much of this focus has been on emerging technologies such as artificial intelligence, nanotechnology, and biotechnology. Similarly, it has long been the contention of science and technology studies (STS) and the social studies of science that transformative technologies, given their vast and pervasive impacts on society, should be framed and shaped by, designed for, and deployed to support democracy (Montes & Goertzel, 2019; Foladori & Invernizzi, 2008; Beumer, 2021). Recent developments in the field of quantum technologies (QT), particularly quantum computing, provide strong reasons to attend to the impacts of that technology also (Coenen & Grunwald, 2017; Vermaas, 2017; Coenen et al., 2022), which is the topic we take up in this paper.

QT are a set of emerging technologies reliant on the utilization of quantum phenomena and separated in different fields such as sensing, communications, simulation, and computation (de Touzalin, et al., 2016). This process of emergence is commonly referred to as *the second quantum revolution* (Dowling & Milburn, 2003) to historically distinguish it from *the first quantum revolution*, which yielded technological artefacts such as transistors, medical imaging devices (like MRI), and lasers. These previous devices created the backbone of technical capabilities which allowed the flourishing of the ICT sector, and yielded products (such as computers and smartphones) that fundamentally altered how several aspects of society function. Though semiconductors used for transistors, and optical elements used for lasers require understanding of quantum mechanics, they do not embody or exploit certain quantum phenomena such as entanglement. Recent developments in technical fields such as nanotechnology, quantum optics, and condensed matter physics resulted in capabilities that allow researchers to reliably create, manipulate, and exploit quantum specific phenomena at a high precision and accuracy levels that were previously not possible. This enabled further research and development efforts that supported *the second quantum revolution*.

The extent of this technological revolution can be observed quantitatively. The number of academic articles, patents granted, and start-ups founded focused on QT has been increasing rapidly in the last decade (Seskir & Aydinoglu, 2021; Seskir & Willoughby, 2022; Seskir, Korkmaz, & Aydinoglu, 2022). As of early 2022, there is nearly 30B Euro of committed or promised public funds for QT until 2030 in terms of national initiatives (QURECA, 2022). In comparison, the literature on 'social' aspects of QT is limited to only two-digit numbers (Wolbring, 2022). Some experts in the field expect that QT could lead to a more unequal distribution of power and wealth between those who are already operating at a very high level technologically and those who do not (de Wolf, 2017, pp. 274-275; DiVincenzo, 2017, p. 248).

Considering this risk, the rapid increase in the academic and commercial landscapes, the relative lack of the literature on the societal aspects of these technologies, and the lagging of public communication and discussion on QT behind the rapidly increasing public and private funding, we argue that a closer examination on how the notion of democratization is being used, and how democratization may work in the field, is necessary.

The democratization of technologies seems to be beneficial *prima facie*, but is however far more complex. There are various schools of thought underlying the meaning and mechanics of what it means to democratize technology, each bringing with them both philosophical as well as technical challenges (Bijker, 2010; Tringham & Lopez, 2001; Verbeek, 2013). In any case, the expertise necessary to innovate in any given technical domain usually accrues in the hands of small groups of experts, thus resisting the basic tenets of democratic politics where all should equally contribute in principle. Still, the huge impacts of transformative technologies demand that how they are designed and deployed incorporate democratic deliberation to achieve more acceptable political decision making.

This paper aims to make a start with filling this lacuna by considering different theories of democracy (participatory, representative, and deliberative), and how the concept of democratization can be formulated with respect to these theories under the postulation of democracy as an intrinsic or instrumental value, without going into broader discussions of ethical or other philosophical aspects. These understandings of democratization are then used to frame how the various actors within the QT community appropriate and use the language of democratization. This is followed by an exploration and analysis of narratives, counter-narratives, and actions concerning the democratization of QT. We focus in our discussions on mainly quantum computing. In the conclusion section, we argue that although the ongoing efforts in the democratization of QT are necessary steps, they should not be accepted as sufficient to say that QT is a democratized field. We argue that more reflexivity and responsiveness regarding the narratives and actions adopted by the actors in the QT field, and making the underlying assumptions of ongoing efforts on democratization of QT explicit, can result in a better technology for society.

## 2. Modalities of Democratization

The notion of democratization in innovation and technology development can be situated in the vast literature concerning the democratization of science in knowledge societies (Domènech, 2017). The issue of how to incorporate democratic deliberative processes into what are expert domains of knowledge has been a longstanding issue in STS with no clear-cut or widely agreed-upon solutions. This has led to a large number of contributions that raise the question of how democratic systems can still function in increasingly complex knowledge societies (Callon et al., 2001; Grunwald et al. 2006; Marres, 2007) particularly given that the rationales regularly given by science managers in public deliberative spaces often distill down to opaque and complex quantitative models of democracy (Tickner & Wright, 2003).

As a potential solution, scholars have proposed that expertise itself serves as the ideal locus for democratization, opening up the possibilities for laypeople to gain expertise and have a better say in our knowledge societies (Lascoumes, 2002). Still, as Miquel Domènech aptly states, such an enterprise even if "taken seriously does not mean it is an easy or even possible task" (Domènech, 2017, p. 127). This, therefore, further raises concerns regarding the logical compatibility between democratic processes and expertise *per se* (Domènech, 2017; Liberatore and Funtowicz, 2003).

These concerns are perhaps even more important in the case of QT, as their relevance and functioning are often more difficult to explain to non-scientists than other technologies.

Various approaches have been proposed and used to democratize science and technology governance in knowledge societies, mostly via programs of stakeholder and citizen participation. This is often understood as the various means by which stakeholders can be involved in the policy cycle which includes:
     1) identifying policy priorities,
     2) drafting the actual policy document,
     3) policy implementation; and
     4) monitoring implementation and evaluation of the policy's impacts
     (OECD, 2020)

Stakeholders are defined differently across the literature. The OECD (2020) defines stakeholders as:

> any interested and/or affected party, including: individuals, regardless of their age, gender, sexual orientation, religious and political affiliations; and institutions and organisations, whether governmental or non-governmental, from civil society, academia, the media or the private sector (OECD, 2020).

Still, there are longstanding and more generalized conceptions of stakeholders within the realm of participatory and value-sensitive design, the latter of which makes a distinction between two classes of stakeholders, i.e., direct and indirect stakeholders:

*Direct Stakeholders*
An individual or group who interacts directly with a technology. For example, a system of electronic medical records might be designed for doctors and insurance companies (Friedman and Hendry, 2019; Friedman et al., 2017).

*Indirect Stakeholders*
An individual or group who is impacted by a technology but does not directly interact with it. For example, many electronic medical systems are intentionally designed to be used exclusively by doctors and, depending on the country, by insurance agencies. This, naturally, will impact patients (Friedman et al., 2017). Another example is when a small drone flies over a bystander, they may be bothered by its sound and presence and their privacy might be violated. In this case, the bystander would be an indirect stakeholder and the operator of the drone would be a direct stakeholder (Vermin et al., 2022).

Still, regardless of the definition(s) used for the stakeholder populations chosen to be elicited, the functionally significant point is that stakeholder groups are identified and given consideration in information, consultation, and engagement processes. Particularly given their role concerning the creation and delivery of policy decisions guiding science and technology innovation rather than in the direct day-to-day technical work of experts.

Although often used interchangeably, *deliberative* and *participatory* democracy have unique elements despite sharing many commonalities. The distinction is important not only for QT designers but for policymakers as well. There is a third conception of democracy, less referred to within the realm of science and technology policy but in general still more relevant than the other two, and that is *representative* democracy.

The number of participants that are included, the type(s) of participation, as well as the method used to select participants differ between deliberative, participatory, and representative democracy theories. For one way of distinguishing them, see Table 1:

|  | Number of participants | Type of participation | Participant selection method |
| --- | --- | --- | --- |
| Deliberative democracy | Relatively small (but representative) groups of people per activity. The aim is to have **deep deliberation**, which is difficult to achieve with a large number of people. | **Deliberation** requires that participants are well-informed about a topic and consider different perspectives in order to arrive at a public judgment (not opinion) about "what can we strongly agree on." | Ideally, a **civic lottery**, which combines **random selection with stratification**, to assemble a public body that is: representative of the public, able to consider perspectives, and not vulnerable to being stacked by representatives of powerful interest groups. |
| Participatory democracy | Large numbers of people, ideally everyone affected by a particular decision. The aim is to achieve breadth. | **More participation**, in all aspects of politics, from all citizens who choose to be involved; an embrace and encouragement of a **diversity** of opportunities for political engagement. | **Self-selected participation** in order to enable as many people as possible to share the experience. |
| Representative democracy | Large numbers of people, ideally everyone affected by a particular decision. The aim is to achieve majority representation. | **Representation** requires representative candidates to be chosen and approved by constituent populations as the voice for their values and concerns. | **Selected candidate participation** determined by the total number of participants as representatives of their values and concerns. |

Table 1. Key differences between deliberative, participatory, and representative democracy. Source: OECD, (2020), based on descriptions in Carson and Elstub (2019) [modified].

In their *Deliberative Democracy Handbook* (2005), Gastil and Levine define deliberative democracy as that which "strengthens citizen voices in governance by including people of all races, classes, ages and geographies in deliberations that directly affect public decisions." A basic idea is that political decision-making processes become more rational and their results more acceptable through the involvement of well-informed citizens and stakeholders in deliberation. This approach to the democratization of science became popular in academic literature during the 1980s with authors like Jürgen Habermas (1981) and Jane J. Mansbridge (1980).

Participatory democracy, on the other hand, has its roots in the civil rights movements that characterized the 1960s (Pateman, 1970). The motivation behind this form of democracy was the demand by citizens, particularly by unrepresented groups, to have greater influence in public decision making processes. Participatory democracy differs from deliberative democracy in that its emphasis is not primarily on how citizens can influence established, usually representative decision making processes from the outside, but on how institutions can augment citizens' abilities and capacities to participate in decision making and make such participation more meaningful (Pateman, 2012).

In representative democracy in the traditional sense, i.e. before the deliberative and participatory turns, deliberation was mostly non-public and limited to elected representatives, invited experts, and rather small numbers of interest groups as only involved stakeholders.. However, this constellation is not typical of most political discourse on science, technology and innovation. Participants in this discourse tend to have less clear roles than participants in the core processes of representative democracy (e.g. in parliament) and may be (even) less representative of society in general and of affected populations in particular than in other policy areas – which is an argument for including more deliberative and participatory mechanisms in the system. (Grunwald et al. 2006; Disch, 2009; Brown, 2009).

In science policy, two major loci for democratization pathways can be distinguished (see Table 2), one on the front end (access to education; science funding) and one on the back end (open access publishing; accountability).

| **Front End** | **Back End** |
| --- | --- |
| Access to Education | Open Access Publications |
| Funding Allocation | Accountability |

Table 2. Modalities of applied democratization in innovation

On the front end, opening up access to education can strengthen democratic knowledge societies. Likewise, democratic (and in particular deliberative and participatory) decision making concerning where funding is allocated can permit a form of public democratic engagement of what spheres of innovation should receive attention.

On the back end, open access publication available without costs to readers (despite the epistemic hurdles in comprehension) is a key endeavour currently being undertaken (see Else, 2021).

Accountability, however, remains a more nebulous and abstract modality for democratization in innovation. Within the Responsible Research and Innovation (RRI) literature, scholars have noted that innovation spheres have mostly remained separated from citizen engagement (Stahl et al., 2021). Owen et al. (2021) argue that this separation is the consequence of innovation spheres not sufficiently aiming to be a 'site for politics', i.e., a domain where deliberation, continuous debate, negotiation, and discussion concerning how innovation should be governed as well as what goals innovation should be driving towards.

The concept of accountability is dealt with in a substantial corpus in the philosophical literature concerning technology. Accountability has traditionally been described as a retrospective, backwards-looking form of responsibility (i.e., van de Poel, 2011) or as a passive form of responsibility, more specifically as after-the-fact evaluation demanding justification which, consequently, provides the foundations for assigning blameworthiness (Pesch, 2015). However, recently scholars have interpreted accountability such that it could have both an anticipatory and preventative conception. This is grounded on an understanding of accountability where there is an active relationship between stakeholders and a forum where the conduct of those stakeholders is uncovered, debated, and justified in continuous public dialogue (Bovens et al., 2014; Santoni de Sio & Mecacci, 2021).

Understanding accountability in this way permits greater synchronicity with normative political theory. In particular, contemporary understanding of representative democracy centers on the importance of accountability as a function of citizens' capacities to self-determination, ensuring that their representatives are responsive to their values, and to ensure that those representatives can be held accountable (Palumbo & Bellamy, 2010).[1] Intrinsic to this understanding of accountability is the notion that the representatives of citizens have a real impact on public decision making processes, and that, hence, these representatives who are responsible for facilitating those processes within institutions, must be sufficiently responsive to the citizens and their values (Schuppert, 2014), and not too late in the process. As a consequence, what should arise to promote public dialogue are deliberative spheres where citizens are encouraged to make their voices heard in decision making processes with at least some connection to the relevant decision-making processes and some interaction between the represented and their representatives (Fung, 2006; Grunwald et al. 2006).

---

[1] Of course, there is no consensus concerning a single understanding of political accountability. For more, see e.g. Bellamy et al., (2011) and Pitkin (1967).

Still, this does not entail that such deliberative spheres are necessarily meaningful. It has thus been argued that participatory spheres should be created where real influence on policies can actually be exerted (Paterman, 1970: pp. 70-71). This, of course can lead to certain asymmetries in participatory spheres where Pareto distributions lead to certain individuals or groups who garner more influence and control to have, similar as in lobbyism, a disproportionate amount of decision making control, thus influencing policy outcomes often to their own benefit (Papadopoulos & Warin, 2007). This, consequently, means that this understanding of accountability may reinforce existing spheres of influence and destabilize the equal opportunity of citizens and stakeholder groups to participate in public decision making concerning science, technology and innovation. However, there has been research that suggests that equitable forms of accountability can be designed into substantive forms of representation to ensure that policy outcomes reflect individuals or groups with less structural influence or power (Page et al., 2013; Grimes & Esaiasson, 2014). These forms of accountability include structural mechanisms to uncover and scrutinize potential conditions of power imbalances where certain agents or groups may cease to reflect the values of stakeholders in decision making arenas (McGeer & Pettit, 2015).

Accountability thus concerns forward-looking governance frameworks, policy choices, spheres, and continuous mechanisms to check undue power asymmetries. If we aim to understand accountability viz. normative political theory, then it becomes *de facto* multidimensional. It involves those who do the accounting and those who are accounted for, it provides the shared guidelines for how behaviors are to be reported, justified, and assessed, and, finally, how violations are to be sanctioned and how those sanctions are to be enforced.[2]

One final caveat to explore here might be the distinction between focusing on the instrumental and intrinsic value of democracy. One might presuppose that democracy as an intrinsic value should always prevail. We argue that this is not the case. Democratic approaches are not merely ideological, they have practical value for their adopters, especially in commercial and industrial contexts. Stakeholder engagement at an early stage might reduce the risk of a product being rejected at launch, hence enabling companies to invest higher amounts or the industrial designers to work more freely on the product. Similarly, widening the base of participation on the development of a technology can attract diverse talent to the field and support innovation processes to explore the section of the opportunity space that was previously inaccessible. Policies developed and implemented by companies to realize these outcomes *de facto* support democratization of a technology in general. However, it might be an open question whether this is acceptable when public policies are enacted where a democratic approach is adopted due to its instrumental value. A for-profit company utilizing deliberative processes to maximize the adoption of its software stack by its potential future user base should in essence be distinguishable from public bodies' attempts of including and informing stakeholders to support the democratic process.

---

[2] For greater depth on the mechanisms of this multi-dimensional understanding of accountability see Buchanan and Keohane (2006).

The layer of accepting democracy as an intrinsic or instrumental value can be thought of as an additional dimension to the distinctions between having different theories of democracy, clear identification of direct and indirect stakeholders, and formulating compatibility between different modalities of action (back end and front end). This can be briefly described as whether utilization of a democratic approach for a situation is assumed due to its instrumental value (i.e., it would yield beneficial outcomes) or accepted as an intrinsic value. Some examples on how the same actions can be formulated with different values is given in Table 3.

|  | Instrumental value | Intrinsic value |
| --- | --- | --- |
| Deliberative democracy | - Including stakeholders to support the technology development and/or adoption<br>- Informing the stakeholders to reduce hesitancy towards the technology due to uncertainty | - Including stakeholders for betterment of the deliberation process<br>- Informing the stakeholders to empower them for them to better present and defend their positions for deliberation |
| Participatory democracy | - Raising awareness to justify public funding<br>- Supporting education efforts for workforce development | - Raising awareness to inform the public and remove the enigma element<br>- Supporting education efforts to distribute expertise among the public and empower citizens |
| Representative democracy | - Aiming for representation from the widest possible majority of stakeholders to support the ecosystem and market formation efforts | - Aiming for representation from the widest possible majority to have most of the public's interests represented |

Table 3. A matrix of value type and democracy theory with several examples of how actions can be formulated

In summary, democracy cannot and should not be viewed as monolithic but rather as multi-faceted with various approaches that can be taken to open-up democratic modes in science and technology research and development. Here we have highlighted three approaches in democracy theory and practice as well as how they are of instrumental and intrinsic value. We have likewise highlighted some modalities of applied democratization in innovation.

In the following section, we will first focus on how the notion of democratization is used in the quantum computing field, and afterwards discuss narratives and actions that support or obstruct efforts of democratization.

# 3. Democratization in Quantum Computing

The democratization of any given domain is not an obvious or straightforward means by which to include important stakeholders in how science and innovation progress. We discussed three democracy theories that may be useful for democratization in the realm of science, technology, and innovation. In the first part of this section, we look at the actual efforts geared towards the democratization of quantum computing. The second and third parts explore the various narratives around democratization of QT more broadly.

Quantum computing promises the ability to make certain calculations possible that supercomputers can practically not perform. For example, it may offer the possibility to simulate (chemical and pharmaceutical) materials much more efficiently, enabling previously not possible research and development avenues to explore, with considerable societal impact (Möller & Vuik, 2017). Considering that the potential uses of quantum computing are yet to be discovered but expected to be societally relevant, some have argued for accessibility as a core value for the field (Coates et al., 2022, p. 9; Coenen, et al., 2022, p.5). In the following part, we introduce efforts by actors in the quantum computing field to realize this, and how it connects to the discussions on democratization.

## 3.1 Efforts in Democratization of Quantum Computing

In May 2016, IBM put the first quantum computer in the cloud (Mandelbaum, 2021) for it to be accessed by anyone interested in this technology. Following this, they announced and developed Qiskit, an open-source software development kit for working with quantum computers at the level of circuits, pulses, and algorithms (Gambetta & Cross, 2018). These efforts by IBM are still ongoing (Wootton et al., 2021).

Briefly, IBM's efforts for education and outreach of quantum computing are:[3]
- First open access quantum computer in 2016.
- Over 100M USD investment into quantum education efforts in the last five years.
- Three million people reached via tools like the Qiskit textbook, YouTube channel, hackathons, summer schools, and so on.
- IBM Quantum Internship accepts and trains 135 students yearly.

These efforts are mentioned in detail here, because this approach is adopted widely for discussing the democratization of quantum computing, which puts a strong emphasis on putting quantum computers in the cloud and providing access to the widest possible range of users, and acknowledged in the literature (Ten Holter, et al., 2022). Similar efforts and approaches to IBM in this field are adopted by others in the community. Quantum computing hardware developers such as D-Wave and Xanadu are also quite actively supporting the open cloud access model. D-Wave has an applications database with more than 250 early quantum applications available.[4] Xanadu has a programmable photonic processor in the cloud that achieved quantum computational advantage (Madsen, et al., 2022). Providing access through the cloud and supporting this access with complementary (such as educational or case-study based industry-oriented) activities can be understood as the main arc of the democratization as it is conceived of in the quantum computing field.

---

[3] Numbers obtained from the "Panel discussion: how do you build a quantum workforce? What is a good way to ascertain what knowledge level and number of experts is required?" This panel was organized on May 18th, 2022 at the Commercialising Quantum 2022 event by The Economist, numbers were given by Liz Durst, Director of the IBM Quantum & Qiskit Community.

Here nuanced distinctions can be made between the uses of the terms 'democratization' and 'democratizing access,' and between the intended target groups. Although sometimes they are used interchangeably (Grossi, 2021), these two terms signify different concepts, as democratizing access can be a part of the democratization efforts, but democratization is a concept that (potentially) encompasses a wider set of actions, including but not limited to *engagement with indirect stakeholders*.

Similarly, some in the field advocate that democratization is for anyone with an internet connection (IonQ, 2022), while others posit it as primarily for developers and researchers (Osaba et al., 2022, p. 55808), and sometimes particularly for non-quantum experts (Liscouski, 2021).

Arguments for the necessity of democratization in quantum computing from a pragmatic point of view usually rely on the phenomenon of workforce or talent shortage (NSTC, 2021; 2022). There are already studies on describing the landscape of the quantum workforce (Kaur & Venegas-Gomez, 2022) or assessing the talent needs of the quantum industry (Hughes et al., 2021). In this regard, a discussion among the community on who needs to be educated in QT, on which topics, to what extent, and for which purpose can also be thought of in its correspondence to the democratization of quantum computing, as introduced above. It is closely related to the distinctions between direct and indirect stakeholders (see Section 2), and whether democratization requires access by only direct or both types of stakeholders, and to the question of whether democratizing access to quantum hardware necessarily entails a reduction of the entry barrier to obtain the required skills to utilize that access or not.

The topic of access to quantum hardware is an analogue of the comparison provided in Table 2 between front end and back end for education. Realization of quantum computers that are commercially active contains the risk of deepening the existing societal divides and inequalities (de Wolf, 2017; Ten Holter et al., 2021). A back end motive would be to have distributed expertise and a widespread understanding to distribute the benefits generated by these devices to the widest possible reach, however, this does not directly translate to front end since different identifications of direct stakeholdership would yield distinctly different actions to be adopted.

---

[4] https://www.dwavesys.com/learn/featured-applications/

Similarly, the trend of leading companies rushing to put their devices freely available on the cloud in an attempt to democratize access can be better understood through them taking democracy as an instrumental value. As listed on table 3, creating a user base helps with reducing hesitancy towards the technology, supports education efforts for workforce development, and aims at increased adoption of the technology to support the ecosystem and market formation efforts. In this regard, companies that manage to reach out to most early users might get strong first mover advantages as the ecosystem and market formation are path-dependent processes. On the other hand, there are many public and grassroots initiatives that are enabled by the actions of these companies. These initiatives aim at raising awareness, inclusion of the widest possible majority, empowering stakeholders, distributing expertise among the public, and much more. Their aims align with taking democracy as an intrinsic value, supporting democracy for the sake of democracy. In the case of companies, their motives are understandably aligned with commercial interests, nonetheless enabling adoption of actions by other actors to further democratize the field.

In the literature, access is not always necessarily associated with the term democratization, and we can clearly see this from the previous literature that advocates for making this technology accessible in one way or another (de Wolf, 2017; Johnson, 2019; Coates et al., 2022; Coenen et al., 2022). As an example, in the Insight Report on Quantum Computing Governance Principles of the World Economic Forum, the stakeholders are listed as governments, academics and universities, international organizations, corporations, private entities that are developing and using the technology, developers, and consumers (Coates et al., 2022, pp. 7-8). In this report, the public is located as an entity that needs to be educated (p. 5) and enlightened (p. 7), and the deliberations are to be held among stakeholders (p. 8; p. 19). The report makes no references to democracy or democratization, but in the taxonomy provided above, one might argue that it is implicitly advocating a model similar to the *deliberative theory of democracy* that limits deliberation to direct stakeholders.

A different argument in the literature without explicitly invoking the terms democracy or democratization is found via the term 'public good' (Roberson et al., 2021). In this literature, authors argue for QT to benefit the societies they will be used in, and maintain that the notion of 'public good' should be extended beyond its traditional conceptualization by economists who describe public goods as those which can be used by many without reducing the availability of said goods. They argue for it to be determined through processes of reasoning and engagement between science and society (p. 3). This requires wider public consultation and engagement if QT should work for a broader societal good. They highlight that current national strategies are formulated in a narrow sense of public good based on themes of increasing national competitiveness and concerns over threats to national security (p. 5), which can be considered as a premature 'imaginary lock-in' (Mikami, 2015) to certain visions by a group of experts focused on realizing a narrow set of futures at the expense of alternative potentialities. This argument can be extended to the limits of social shaping of QT, where a lack of public clarity regarding concrete expectations for QT prevents specific engagement around societal impact (Roberson, 2021a, p. 394). Premature lock-in to certain visions via national narratives limit wider public consultation and engagement, hence limiting the impact of social forces on shaping the trajectory of QT. Finally, it is discussed that for a responsible development of QT, increasing public awareness is the absolute minimum that needs to happen (Coenen & Grunwald 2017; Roberson, 2021b). These points in the literature signal that there needs to be further discussion on public engagement and democratization of QT, even if in the end society just endorses the prematurely locked-in visions by national strategies. This approach is more in alignment with *participatory democracy* in the taxonomy given in Section 2, where the whole of society (whether they are direct or indirect stakeholders) should be included in the process of guiding the development trajectory of QT.

A third overall tendency that can be observed in the quantum community and previously mentioned in the literature is through community membership, advocacy, and "the maker movement" (de Wolf, 2017, p. 275), which can be associated with the *representative theory of democracy*, but one that is enriched by deliberative and participatory approaches. This is mainly enabled by public investment programs (like the Quantum Community Network (QCN) under the EU Quantum Flagship[5] and the Quantum Future Academy[6]), companies that are developing quantum hardware and software resources (spearheaded by IBM with the Qiskit Advocate initiative[7]), by the formation of online global grassroots communities[8] (such as QWorld, OneQuantum, Full-Stack Quantum Computation, and so on) that lowers the entry barriers to the field of quantum computing, and by communities focusing on special interest groups[9] (such as Q-munity for high school students, Girls in Quantum for girls and students, Women in Quantum Development (WIQD) for female professionals in QT, Q-Turn (Sainz, 2022) for researchers from marginalized groups, and so on). These efforts do not mainly discriminate between direct or indirect stakeholders and promote the widest possible representation in the QT space. QCN has leading researchers representing their countries, Quantum Future Academy hosts two students from each countries selected by national scientific institutions, Qiskit Advocates act as mentors and liaisons between enthusiasts and IBM's resources, grassroot communities such as QWorld and OneQuantum have local branches in more than 25 countries with *de facto* national representatives of the local quantum grassroots communities, and communities such as WIQD and Q-Turn aim to give voice to researchers representing certain interest groups. All these efforts can be viewed in alignment with a *representative theory of democracy after the deliberative and participatory turns*, where different stakeholders and populations contribute and become a part of how *the second quantum revolution* unfolds via certain representatives acting as intermediaries.

---

[5] https://qt.eu/about-quantum-flagship/the-quantum-flagship-community/quantum-community-network/
[6] https://www.quantentechnologien.de/bildung/quantum-futur-akademie.html
[7] https://qiskit.org/advocates/
[8] https://qworld.net/, https://onequantum.org/, https://fullstackquantumcomputation.tech/
[9] https://www.qmunity.tech/, https://girlsinquantum.com/, https://www.wiqd.nl/, https://www.q-turn.org/

As listed above, approaches in line with one or more of the three theories of democracies can be seen in both the literature and the practice of quantum computing, with a particular focus on awareness raising and education due to the phenomenon of workforce or talent shortage. In the next subsection, we will focus on narratives and actions against democratization in quantum computing and QT more generally.

## 3.2 Narratives and Actions Against Democratization

In the current QT ecosystem, we observe three major obstacles for opening up QT to a wider public engagement process that also incorporates a diverse set of perspectives from different groups of stakeholders. These are the narratives of, (i) QT as an arena for geopolitics, (ii) quantum mechanics as incomprehensible, and (iii) quantum computing as a threat to the cyber-infrastructure. In this subsection, we first describe and analyze these narratives and related actions. As a next step, we will discuss the counter-narratives and actions against these obstacles.

*QT as an Arena for Geopolitics*

The first narrative is the one of QT as an arena for geopolitics, and its discussion takes up the larger part of this subsection. It is widely accepted that there is a *quantum arms race* going on (Giles, 2019), mainly between the US and China, but encompassing a wide range of alliances. Reflections of this line of thinking can be seen especially regarding international collaboration. Quantum computing and QT in general have been a topic of geopolitics for some time. Export controls under the Wassenaar Arrangement are implemented for some of the QT (Bureau of Industry and Security, 2019), there are special arrangements and limitations on cooperation between certain countries (Ten Holter, et al., 2022), and it has been discussed previously that the global cooperation between rivaling powers was fizzling out (Biamonte, Dorozhkin, & Zacharov, 2019) even before the SARS-CoV-2 pandemic and the invasion of Ukraine. Some might argue that there are good reasons for that. Certain application areas within QT are in direct alignment with military and defense industries, such as positioning, navigation and timing (PNT), cryptography, and high performance computing (HPC). Furthermore, historically (especially in the United States), new technologies have been funded by military efforts first and civilian applications followed later. Finally, a kind of vigilant technology sovereignty argument is accepted as a legitimate point to argue for public funds in an increasingly tensioned global political stage that brings together the technological superiority narrative with the "Might makes right" mindset. For a recent example, the newly formed AUKUS, a trilateral security pact between Australia, the United Kingdom, and the United States, has QT under its advanced capabilities areas for cooperation and an arrangement titled "The AUKUS Quantum Arrangement (AQuA)" is set up to accelerate investments to deliver generation-after-next quantum capabilities (The White House, 2022). These technologies are sometimes presented as "transformational technologies" or "game-changing advances" (McKay, 2022, p. 6) for the military branches such as the US Air Force. This is mirrored by efforts in China, where, following the Snowden Revelations, a considerable effort to 'hack-proof' their critical infrastructure was initiated (Chen, 2014). However, it has been shown that research on quantum cryptography in China preceded 2013, and China has been a leading actor in scientific efforts on quantum cryptography, at least quantitatively, since 2007 (Olijnyk, 2018). This can also be observed in terms of patents, as China has become the country with most patents granted in QT (Seskir & Willoughby, 2022) in a decade via an aggressive patenting policy. A similar stance can also be observed in the European Union, as phrases like "Ensuring security and technological sovereignty" (Castelein, Ormanin, & Kostka, 2020, p. 37) or "...reaching its political ambition in Quantum Technologies, in order to safeguard European strategic assets, interests, and security…" (Quantum Flagship, 2022) can apparently be encountered more often than before. So, overall, QT is presented as an arena for geopolitics by some actors of that arena and of the QT community.

These points enable national actors to prioritize certain militarized visions of QT against others, even before they are brought forth to the public arena for discussion, and this poses a major obstacle in democratization of QT in four different respects. First, it is creating a reasonable ground for diversion of research funds to organizations with security priorities rather than scientific ones. One concrete example of this is the work on quantum radar (Lanzagorta, 2011), which relies on a technique called quantum illumination (Lloyd, 2008), where most of the research is conducted in organizations either directly working with militaries or in close association with them (Durak, Seskir & Rami, 2022). News articles starting with sentences such as "A new quantum radar technology developed by Chinese scientists could detect stealth aircraft…" (Chen, 2021) are part of a discourse that reinforces the militarized priorities and strengthens the mode of thinking that supports restricting access to knowledge, instead of increasing it.

The second point is the exclusion of students and researchers from certain backgrounds. In recent years, there were discussions over the FBI visiting American universities "...with an unclassified list of Chinese research institutions and companies" (Feng, 2018), asking them to monitor students and researchers from these institutions and companies. Although calls for keeping quantum computing research open were being made (Biamonte, Dorozhkin, & Zacharov, 2019), now the field is becoming more and more balkanized. This has direct consequences for democratization efforts. For example, certain grassroots communities that operate in the EU and the US ceased their operations in Russia and China in the last few years and some do not even accept participants to their otherwise globally open events from these regions. Companies, especially start-ups, sometimes refrain from collaborating with partners that also have or had collaborations in these countries. Furthermore, it is sometimes openly discussed that access to not just devices, but also the knowledge networks should be restricted, not only to researchers in Russia or China, but to any researcher that is operating outside of the states governed by 'like-minded governments'.

The third point is the exclusion of researchers that for one reason or another prefer not to work in alignment with military objectives on certain research topics. There are statements in the literature on "Should QST support war?" (McKay, 2022) or even "Should We Build Quantum Computers at All?" (Chen, 2022). There are active members of certain quantum communities referring back to their unknowing involvement in defense-related projects in QT during their graduate studies and how it actually drove them to dropping out of academia. There are some departmental customs in many academic institutions (especially in social sciences) that discourage researchers from getting involved in defense-related topics. As QT becomes more militarized, it may repel researchers with a focus on social justice and then evade scrutiny by those that may otherwise propose interventions on the trajectory QT is progressing in.

The final point regarding the first narrative is the reduction in awareness-raising and outreach activities as certain parts of QT become militarized. As militarization of a technology excludes certain researchers from entering the research domain, it also limits the amount of information that is getting out of the research labs. Researchers involved in defense-related projects may do less outreach, collaborate more internally, and rarely engage in citizen science activities – regardless of the question if the researched technology will ever be field-deployable.

A counterargument to the points we presented above can also be developed. One could comment that the geopolitics around QT may in some cases actually defend democratic values. By developing QT autonomously specifically democratic nations can defend their technological sovereignty and offer its citizens democracy as well as democratic control over these technologies. When considering this possible counterargument with the three democratization theories introduced in Section 2, it can be noted that this offered control is at max democratic control at the national level and in the sense of deliberative democracy. By excluding citizens of specific other nations, an international democratic control of QT is frustrated, and by shifting research to the military domain with the accompanying reduction of transparency about this research, large groups of citizens are effectively not enabled to exercise meaningful control, as is envisaged in participatory and representative democracy.

*Quantum Mechanics as Incomprehensible*

The second narrative is about quantum mechanics being incomprehensible. Richard Feynman famously wrote, "I think I can safely say that nobody understands quantum mechanics" (1995, p. 129). One iteration of this quote is used in a 2013 article by the science writer Philip Ball at BBC titled "Will we ever… understand quantum theory?", citing a survey study on foundational attitudes toward quantum mechanics (Schlosshauer, Kofler, and Zeilinger, 2013). Ball reassures his audience that if "the baffling behavior of subatomic particles leaves you scratching your head with confusion, don't worry. Physicists don't really comprehend it either."

This is just a single example out of a myriad of popular science articles, news pieces, and presentations for the purpose of education or outreach on quantum mechanics; so much so that it is a common practice among physicists giving public talks to quote Feynman, saying that quantum mechanics is not understandable, and then try to make their audience understand quantum mechanics. This issue gets further complicated as promoters of pseudoscientific ideas such as quantum healing (Chopra, 1990), quantum mind, and quantum consciousness utilize this approach. The quote and the argument behind "nobody understands quantum mechanics" plays into their narrative: where nobody understands something, the line between science and pseudoscience gets blurred.

Examining this narrative with the lens of the three democracy theories, it is relatively straightforward to observe that it limits legitimate participation into the development process of QT to a small group of experts that somehow manage to make the technology work, even though they also do not fully understand it. It thus effectively blocks democratization efforts in the sense of the participatory and representative theories, and does so even for the future by declaring that citizens cannot ever be in the position to meaningfully decide about QT. It thus practically makes public deliberation about QT impossible under all three democracy theories, leaving us again with limited deliberative democratic control by small expert groups.

*Quantum Computing as a Threat*

The final narrative is quantum computing as a threat. In 2014, following the Snowden Revelations of 2013, an article came out in the Washington Post titled "NSA seeks to build quantum computers that could crack most types of encryption" (Rich & Gellman, 2014). This article gives an overview of the efforts by The National Security Agency of the United States Department of Defense, that covers ominously named projects such as "Penetrating Hard Targets" and "Owning the Net." In 2015, the Information Assurance Directorate (IAD) under the NSA and the Central Security Service (CSS) declared that no commercial security algorithm suite is considered secure in the long run anymore by the IAD. In a document titled "Commercial National Security Algorithm Suite and Quantum Computing FAQ," (2016) they stated that auxiliary measures should be taken "...while waiting for quantum resistant algorithms and protocol usage to be standardized" (p. 5). This is due to Shor's algorithm (1997), which is one of the few quantum algorithms that can prove an exponential speed up for a mathematical problem against its best-known classical counterparts. The mathematical problem in question for Shor's algorithm is prime factorization. It is part of several essential cryptographic algorithms that are widely used for asymmetric public-key cryptography, which secures online transactions between different parties for many purposes such as messaging, sending images, sharing sensitive data (like credit card information for online shopping), and so on. This revelation fueled a narrative that was already there: quantum computing is a threat to the entire cybersecurity infrastructure of the internet as we know it.

A report published by the Global Risk Institute (Mosca & Piani, 2022) containing the results of a survey conducted with over 46 experts in the field shows that the majority of respondents estimate that a quantum computer able to break RSA-2048 in 24 hours is not going to be around in ten years, and they deem it only partially likely in 15 years. According to the survey the real threat horizon begins only after 20 years. Furthermore, these estimates are for RSA-2048, which is highly unlikely to be still used in 20 years. Still, it can be rightfully argued that breaking these algorithms can cause serious problems to data privacy and particularly pose a danger against sensitive data that requires storage for a long time (20+ years). Hence, the final obstacle, portrayal of quantum computing as a threat is a narrative frequently encountered.

This narrative both feeds into the first narrative of QT being an arena for geopolitics and it frames the technology as dangerous, causing protectionist impulses (Ten Holter, et al., 2022) by certain actors to limit the distribution of even much smaller quantum computers that in no way pose a threat by such means as the Wassenaar Arrangement. This framing of quantum computers as potential weapons that can lay waste to the infrastructure of the internet, which is almost an essential utility for the modern society to function, triggers questions such as "Should We Build Quantum Computers at All?" (Chen, 2022). The answer given to this is usually along the lines of "X is already building it, they will build it even if we don't," where this X changes depending on who you ask. This narrative acts as a means to frame the technology in a certain manner that justifies limiting participatory and deliberative processes, nudging the representation space to a more securitized configuration where direct ownership is allowed only to a selected few. Posing quantum computers as a threat is the epitome of creating a justification for restricting access, especially as the technology matures.

Similar to the QT as an arena for a geopolitics narrative, the quantum computing as a threat narrative also supports the introduction of exclusionary mechanisms to the participatory efforts. It creates time pressure on the potential deliberation process and prioritizes certain (security-oriented) stakeholders' needs over others.

Until now in this section, we presented three of the narratives that act as inhibitors for an inclusive public engagement process, which are (i) QT as an arena for geopolitics, (ii) quantum mechanics as incomprehensible, and (iii) quantum computing as a threat. There are other obstacles and narratives that can also be formulated as potentially inhibiting the democratization process. We chose these three since they were already pointed out in one form or another in recent literature (Coenen, et al., 2022; McKay, 2022; Ten Holter, et al., 2022), and often are a topic of informal conversations or conference talks within the QT community. In the following part, we present some alternative points and actions that can be adopted instead of these narratives.

## 3.3 Counter-narratives and Actions Supporting Democratization

There are narratives and actions in the QT community that counter the three narratives described in the previous subsection. First, regarding the narrative representing QT as an arena for geopolitics, a militarized technology is almost by definition not a democratized one. Under any of the three democracy theories, excluding the majority of the public and enabling a selected group access to the operational capabilities of a certain set of technologies, cannot be argued for without some serious supporting situational conditions. Having said that, a democratic technology can still be utilized for military and defense purposes; it may be less effective (since it does not provide a similar edge due to exclusivity) but it is still possible. Furthermore, discussions on how to balance the risks and responsibilities of stakeholders in this domain can be found in the literature (Roberson, 2021b).

For an inclusive process, it should be acknowledged that the world consists not only of hegemonic powers, and QT has the potential for much more than just being another arena for technological competition between them. Civilian uses of QT should be emphasized and how they can be utilized in a way that benefits the common good (Coates et al., 2022, p. 9), such as via the Sustainable Development Goals (Coates et al., 2022, p. 16). Civilian uses should be given primacy over creating benefits to operational capabilities of certain military branches. Furthermore, the most coherent response to the misuses of these technologies can be better formulated with international agreements and other forms of stakeholder action (Silbert, 2022), rather than with siloed approaches in different global camps. Finally, there are already calls in the literature that "...a responsible innovation mindset requires us to think in potentially more creative ways about national approaches, ownership of such technologies and protectionist or nationalistic approaches" (Ten Holter, et al., 2022).

Second, regarding the narrative on quantum mechanics being incomprehensible, counter-narratives and actions exist too. As noted earlier in this section, IBM alone spent more than 100M USD on quantum education efforts in the last five years. There is a committed section of the Quantum Flagship focused on QT education,[10] research institutes reach out to the public at large to make QT understandable,[11] and there are many companies and communities focused on education efforts in QT. There are many innovative approaches such as utilizing games and interactive tools/textbooks (Wootton, et al., 2021; Seskir, et al., 2022), considering quantum mechanics as a generalized probability theory (Aaronson, 2013) and organizing community-based workshops (Salehi, Seskir, & Tepe, 2022). All these efforts rely on the assumption that one does not need to be a seasoned physicist to understand quantum mechanics on a level to be operationalized for the purposes of utilizing QT. Targeted communication efforts concerning pseudo-scientific contributions to public discourse on QT may also be useful.

Interpretations of quantum mechanics is a fascinating topic, and the foundations of quantum mechanics are full of surprising and sometimes confounding ideas and discoveries. This does not mean, however, that quantum mechanics is incomprehensible, and it surely does not support the Feynmanian position that "nobody understands quantum mechanics." For a technology to operate and be adopted by the public, it needs to be understandable, it needs to be 'normal' in a Kuhnian sense. Although one might find Feynman's quote interesting or even funny and aims to use it to instill some sense of wonder and mystery, it is working against the basic assumption that motivates all the outreach and education efforts. It raises the entry barrier to newcomers, it might encourage some that are particularly tuned to challenging this quote, but this will only be a small minority of the public. For an inclusive public engagement process, this narrative should go, and a new one that portrays QT as more normal and mundane be developed: one that is open to participation from all parts of society, one that can be understood through getting involved, and one that is based on science, not mysticism. Interpretations of quantum mechanics can play a role in making QT comprehensible (Vermaas, 2017) as they are descriptions of what the world would be like if quantum mechanics is true. In fact, one could even argue that the descriptions of atoms and qubits that are to be used by engineers working in QT for making these technologies more comprehensible can help favor a specific engineering interpretation of quantum mechanics (Vermaas, 2005).

---

[10] Quantum Technology Education: https://qtedu.eu/

[11] TU Delft has vision teams projects that create dialogues with society about perspectives on quantum technologies (Vermaas et al. 2019) and https://www.tudelft.nl/en/about-tu-delft/strategy/vision-teams. And other institutions (such as KIT) are creating dedicated offices to make QT more understandable and knowledge on the topic more accessible by the public.

Finally, regarding quantum computing as a threat narrative, in 2016 NIST announced that it will be collecting nominations for public-key post-quantum cryptographic algorithms. For over the last five years, a period of open testing has been going for over more than 80 algorithms that were initially submitted. Currently, in Round 3, there are seven main and eight alternate candidates. Even though the process is not finalized yet, some cyber infrastructures already started adopting post-quantum algorithms (such as OpenSSH 9.0 adopting NTRU Prime algorithm[12]). This does not mean that the post-quantum algorithms are entirely secure, it is an active research area and recent research reveals some vulnerabilities in some algorithms that are considered as finalists by the NIST (Karabulut & Aysu, 2021), requiring the algorithms to be updated. Similarly, the European Telecommunications Standards Institute (ETSI) is working on migration strategies and recommendations for quantum-safe schemes (Antipolis, 2020). All in all, it can be safely said that the issue of 'quantum computing as a threat' is being taken seriously, and several institutions around the globe are working on providing solutions, guidelines, and strategies to prevent negative scenarios.

To sum up: We have identified three narratives as major obstacles inhibiting an inclusive democratization process on QT and provided some possible ways to remove these obstacles and alternative narratives to replace the current ones (see Table 4). Education efforts to make quantum mechanics and QT understandable to all may be taken as democratization under all three theories of democracy; the other mentioned counter-narratives and efforts may be directed to and enable only a limited group of stakeholders, fitting more deliberative democracy. The reasons these narratives are widely adopted in the community are more complex than the explanations that we provided. Of course, the alternative narratives we offered are not the only possible ones, and we do not argue that they are necessarily the best alternatives either, but they can act as starting points for more constructive discussions.

---

[12] Patch notes for the release: https://www.openssh.com/txt/release-9.0

| Narratives | Alternatives |
|---|---|
| QT as an arena for geopolitics | - accept that the world consists of more than just hegemonic powers, avoid protectionisms<br>- focus on civilian uses related to sustainable development goals |
| quantum mechanics as incomprehensible | - Feynman's quote should not be worked with anymore, people can understand quantum mechanics<br>- QT needs to become a 'normal' (and even mundane) technology, pseudoscience should be criticized |
| quantum computing as a threat | - focus on ways to constructively deal with threats (such as the NIST process)<br>- present realistic timelines, not only the worst-case scenarios |

Table 4. Narratives as obstacles for inclusive public engagement and some alternatives

In the following section, we expand the discussion to explore the nuances of democratization regarding different QT, how public participation already is and further can be included in this process of democratization.

# 4. Toward More Democratization in QT

Many of the efforts and narratives we discussed so far may support democratization of QT. In this part, we highlight two main lines of discussion: the topics of access and participation. Finally, we ask what else can be done and how shifting focus to more overarching points might provide further insights.

## 4.1 Enabling Access

In Section 3 we pointed out that enabling access to quantum computing hardware is associated with democratization. Several authors have argued that such open access is essential for quantum computing (de Wolf, 2017; Ten Holter et al., 2021; Kop, 2021; Coates et al., 2022; Coenen et al., 2022). For the particular case of quantum computers this makes sense – however, QT is more than quantum computing.

For quantum computing one could argue that giving general access is practically and economically feasible since it can be organized by existing cloud infrastructure. For the other QT providing access may prove to be much more difficult. Consider, for instance, democratizing access to quantum key distribution (QKD) devices. Current costs of QKD devices are around 100k€ for reasonable systems that can handle modest key rates (KEEQuant, 2021, p. 8). Even assuming a hundred-fold reduction in cost, providing access to QKD for all the households in the EU would cost approximately 200B€. One might argue that democratization in this case means giving access to QKD devices by avoiding legislation that would limit public use of QKD. But arguing for widespread distribution of this technology for democratization leads nowhere due to the simple reason that the required funds do not exist and will not exist for the foreseeable future.

A different discussion can be made for quantum sensors. Access to classical sensors (such as gyroscopes, cameras, ranging devices) are common, each smartphone contains a considerable number of sensing tools. This does not mean that users are well-educated on how sensors work, most are even unaware that they have these devices in their smartphones. It was noted in a public dialogue report published by the Engineering and Physical Sciences Research Council of the UK (2018) that participants were "...confused, disengaged or neutral and communicated finding the topic difficult to understand" (p. 8) when the inner workings of gravity sensors were presented to them, but they responded more positively after learning that such devices can be used for climate monitoring (p. 26). This again brings forward the question of what exactly should be democratized when such complex and layered technologies are in question: access to the underlying mechanics or the products utilizing those low-level physical properties for operations that are more recognizable by societal actors and stakeholders. Deliberative processes with stakeholders and citizens more broadly on applications of QT thus may be particularly effective democratization measures.

Another interesting aspect is the democratization of research tools and results within the QT community. To quote one example concerning research on qubits, the basic key components in quantum computers:

> "There are other ways to democratize spin qubits. One can distribute known good devices to academic groups interested in exploring new qubit encodings and ways to control them; this is the foundry model. Increasing throughput of testing at multiple temperature stages (room temperature, 1 to 4K depending on the physics, and <100 mK) also directly benefits fabricated device optimization." (Tahan, 2021, p. 3)

Although it is not framed in this manner in the given example, this is a particularly valid and valuable discussion as the high costs of devices limit the participation of researchers from the Global South. Increasing the number of facilities globally has practical benefits for research and development purposes as well. It enables access to local talent, it gives access to some national funding as included countries are most likely to divert funds into the topic, and it promotes the research topic in further regions. Regardless of accepting democracy as an instrumental or intrinsic value, distribution of research among international academic groups, and bringing developing nations onboard has some merit that needs to be taken seriously.

To sum up, whether access is necessary or sufficient for democratization of QT does not have a straightforward answer. If we accept the *participatory* depiction of democracy, by not having access, a huge portion of society loses even the chance of self-selected participation. On the other hand, in a *representative* understanding of democracy, the sections that do not have direct access to these technologies can influence the development trajectory of them via representative means, for example, by supporting particular funding schemes. Furthermore, public participation and dialogue exercises, such as the one that was run by EPSRC (2018), are a great example of how a *deliberative* approach can be taken even when the public does not have direct access to these technologies. This approach may be particularly relevant when it comes to an inclusive democratic approach concerning potential applications of QT.

## 4.2 Enabling Participation

Opportunities for public participation in the second quantum revolution is another important avenue for discussion on democratization of QT. The obvious avenue of participation for citizens is as consumers, but even that is not very straightforward, as hinted at when discussing public access to QKD. The near-term QT market is expected to be mainly B2B (business-to-business), meaning that the public will not even be participating as consumers. This means that by the time these technologies are in the market for the individual consumers, there might be little room for them to have any say on the development trajectory due to all the path-dependencies infused into their design.

There are some options explored in the QT community to circumvent this outcome. The first one is forming lobbying groups that represent the public's interests (EPSRC, 2018, p. 43). The second is through participation in research activities via citizen science initiatives. Some examples of this can already be found in the literature (Lieberoth et al., 2014; Heck et al., 2018; Jensen et al., 2021). The third is participating and forming grassroots communities (such as QWorld, OneQuantum, Full-Stack Quantum Computation, Q-munity, and so on), and more specialized communities representing specific groups such as women[13] (WIQD, Women in Quantum, Womanium Quantum, etc.), researchers that are under-appreciated due to belonging disadvantaged communities[14] (Q-Turn), students from certain regions[15] (such as PushQuantum from Munich), quantum software engineers working on open source software[16] (Quantum Open Source Foundation, Unitary Fund, etc.), researchers working on combining quantum and climate sciences[17] (Q4Climate), and many others. Fourth is the companies forming such communities for outreach and education efforts (like the Qiskit Advocate program). Finally, there is a developing culture of publishing 'manifestos' in the European QT community.

---

[13] https://www.wiqd.nl/, https://onequantum.org/women-in-quantum/, https://www.womanium.org/Quantum/Computing
[14] https://www.q-turn.org/
[15] https://www.pushquantum.tech/
[16] https://qosf.org/, https://unitary.fund/
[17] https://q4climate.github.io/

One of the first and arguably the most influential manifesto in the community was published in 2016 (de Touzalin, et al.). Both the manifesto and the list of 3,680 endorsers are publicly accessible:[18]

> "On invitation of Mr. Günther Oettinger, Commissioner for Digital Economy and Society and Mr. Henk Kamp, Minister of Economic Affairs in The Netherlands, a European team has been working on a "Quantum Manifesto" to formulate a common strategy for Europe to stay at the front of the second Quantum Revolution. The Manifesto will be officially released on 17-18 May 2016 at the Quantum Europe Conference that The Netherlands is organizing in Amsterdam in cooperation with the European Commission and the QuTech center in Delft."

This was followed by several other manifestos open for endorsement such as the Quantum Software Manifesto (Ambainis, et al., 2018), The Talavera Manifesto for Quantum Software Engineering and Programming (Piattini, et al., 2020), and the Quantum Energy Initiative (Auffèves, 2022) with varying numbers of signatories, usually around a few hundreds, and by a manifesto that did not look for endorsement and focused on ethical and societal issues (Coenen, et al., 2022).

The different modes of participation also differ in their correspondence in how they relate to democratic representation. Grassroot communities and lobbying groups enable enthusiasts and the members as stakeholders to either elect *de facto* representatives of their interests within the given domain of public debate or create participatory spaces where the democratic process takes place on a smaller scale. The route of publishing manifestos calls for deliberation on several aspects. For example, the formation of the Quantum Energy Initiative (Auffèves, 2022) raises questions on whether energy efficiency should be endorsed as a core design value for developing QT, and if so, what does this entail? Furthermore, there are proposals for sets of 'core values' of QT such as accountability, inclusiveness, equitability, non-maleficence, accessibility, and transparency (Coates, et al., 2022, p. 9) or comprehensible, specific, open, accessible, responsible, culturally embedded, and meaningful QT (Coenen, et al., 2022, p.5) that are vocalized via these manifestos and similar publications. All in all, it is a practice emerged in (mainly) the European QT community that is directly in alignment with the process of democratization in QT.

## 4.3 What else can be done?

As a first step towards furthering democratic participation, there must be an initial recognition that the public consists not merely of uneducated and unenlightened people but is a collection of different stakeholder groups with (sometimes) divergent interests. This makes public participation and engagement complicated but it is still necessary. Providing access to information, education, and actual hardware in some particular cases are good practices, but public participation in the democratic sense requires adoption of a "strong" RRI approach which entails linking parliamentary or other core policy processes (Coenen & Grunwald, 2017, p. 277) to stakeholder dialogues, decision-supporting public engagement and a wide variety of other public communication activities. While this follows a deliberative democracy approach, it is beneficial regardless of the theory of democracy one adopts, given that democracy is taken as an intrinsic value and not utilized for instrumental purposes. Furthermore, other practical steps can also be taken such as supporting the co-creation, with stakeholder groups, of use cases that benefit societal goals, democratization of not only the technologies but visions as well through co-development of cultural artefacts (such as games, artworks, tv shows), opening up controversial topics to discussion – for example, the militarization of QT (McKay, 2022) or the 'supremacy' debacle (Preskill, 2019) – and so on.

---

[18] http://qurope.eu/manifesto

It should be noted that democratization of QT does not necessarily enable the same tools and approaches that work for quantum computing. QT contains a wide range of technologies that are put under the umbrella of QT due to which physical phenomena they utilize. In this sense, we argue that distancing the topic of democratization from particular technological capabilities (such as cloud access) to more overarching points of discussion (such as stakeholder engagement) would benefit the community and societies further.

# 5. Discussion and Conclusion

As QT emerges, it becomes societally more relevant. Similar to the discussions on democratizing nanotechnology (Toumey, 2011), democratizing QT is not a straightforward topic. In this article, we provided some examples of how the concept is used in the community, especially for quantum computing. We argued that the term democratization should be located in the context of theories of democracy, also with a view to the questions whether democracy is accepted as an instrumental or intrinsic value and how direct and indirect stakeholders are identified. Furthermore, we analyzed certain narratives and actions on QT that are obstacles to democratization efforts, together with counter-narratives and actions in support of democratization efforts. Finally, we introduced several practices in the QT community that are working toward more democratization in QT.

We argued that there is practical value in making such conceptual frameworks explicit. First, having a clearer understanding of the context in which the term 'democratization' is located and used allows both, developers and practitioners of policy actions (public institutions, companies, NGOs, etc.) a ground on which they can formulate their aims, goals, and visions. Second, this can help translation of best practices between different technologies, programs, countries, and so on. Some of these contexts are incompatible, such as using a participatory democratic model with interest in its instrumental value versus deliberative democratic model that prioritizes the intrinsic value of democracy itself. Having a clearer understanding of which policy or program adopts what kind of an approach reduces the chance of confusion and helps not only to develop novel policies but also with respect to their implementation and assessment. It should be accepted that different actors have different roles and priorities. It is more likely for commercial actors to use democratization for its instrumental value[19], while public actors aim for policies that promote democratization itself. A model that allows one to dissect the policies into their conceptual cores and to identify the choices taken by the actors in each step can enable a richer analysis rather than just focusing on some bland indicators.

Similarly, analyzing narratives, counter-narratives, and actions by actors in the QT community in their relation to democratization efforts can also benefit from elucidation. For example, the use of certain narratives such as 'quantum computing as a threat' or 'quantum mechanics being incomprehensible' are generally not the message that most outreach efforts aim for[20]. Hence, we hope that as they are made explicit, it will become a standard to replace these narratives with more constructive ones that focus on solutions and realistic timelines against the quantum threat, and on normalization of quantum mechanics and QT to be more inclusive, not incomprehensible and only understandable by a small group of highly trained experts.

Finally, we argue that although there are some commendable initiatives within the QT community, it is not possible to call QT a democratized field. Access to either the technology or the products of it is a necessary but not sufficient condition, hence access does not guarantee agency. Efforts in identification and engagement with not only direct but also indirect stakeholders are rare (Vermaas, 2017; EPSRC, 2018). Furthermore, there are narratives that act as inhibitors for an inclusive public engagement process, which act as exclusionary mechanisms. Considering that we are still in the early days of QT, having a more democratized field compared to previous emerging technologies is possible, but it requires reflexivity and responsiveness by the community, which may provide us "a chance to repair classic(al) mistakes" (Ten Holter, et al., 2022).

In this paper, we purposefully omitted discussing the issue of ethics in QT, which would have further complicated the topic of different theories of democracy in the context of democratization. We acknowledge that there is a sprouting literature (Kop, 2021; Perrier, 2021a, 2021b; Meyer et al., 2022) and a series of community-based efforts[21] on the topic and would encourage the interested readers to get involved with that fascinating and important topic. Similarly, literature on governance of QT has been emerging (Johnson, 2019; Coates et al., 2022), even intersecting with discussions on democratization (Kop, 2021). Exploration of the relations between these concepts are intriguing avenues of research for further studies.

---

[19] An example from Microsoft: "…bringing those innovators to the table and democratizing quantum so we can get solutions out quickly." (Commissariat, 2022)

[20] The proposal to not use the Feynman quote and argument of quantum mechanics being incomprehensible received particularly positively by practitioners of outreach in QT whenever we presented this paper during preparation process.

[21] https://thequantuminsider.com/2021/02/01/quantum-ethics-a-call-to-action/, https://quantumethicsproject.org/, https://qcethics.org/

It has been argued that "society is promised new (quantum) technologies" and that "it will get some, although they will not be things that will be understood by the broad midsection of society" (DiVincenzo, 2017, p. 248). We believe that taking a step back and investigating how connecting the promises of QT with this broad midsection can yield valuable insights. QT has a role in the future of our societies, but it is up to all of us to deliberate on what that role is.